\documentclass[prl,prc,twocolumn,superscriptaddress,showpacs,floatfix]{revtex4}

\usepackage{graphicx,color}
\usepackage{dcolumn}
\usepackage{amsmath,tabularx}
\usepackage[ps2pdf,letterpaper,colorlinks,linkcolor=darkred,citecolor=darkgreen]{hyperref}

\definecolor{darkgreen}{rgb}{0,0.25,0.05}
\definecolor{darkred}{rgb}{0.3,0,0.03}

\newcommand\NSU{Norfolk State University, Norfolk, Virginia 23504}
\newcommand\UVa{University of Virginia, Charlottesville, Virginia 22903}
\newcommand\Jlab{Thomas Jefferson National Accelerator Facility,
  Newport News, Virginia 23606 }
\newcommand\UBasel{Universit\"{a}t Basel, CH-4056 Basel, Switzerland}
\newcommand\FIU{Florida International University, Miami, Florida 33199}
\newcommand\HamptonU{Hampton University, Hampton, Virginia 23668}
\newcommand\MissSU{Mississippi State University, Mississippi State,
  Mississippi 39762}
\newcommand\NCAT{North Carolina A\&T State University, Greensboro,
  North Carolina 27411}
\newcommand\ODU{Old Dominion University, Norfolk, Virginia 23529}
\newcommand\SUNO{Southern University at New Orleans, New Orleans,
  Louisiana 70126}
\newcommand\TelAviv{Tel Aviv University, Tel Aviv, 69978 Israel}
\newcommand\UMD{University of Maryland, College Park, Maryland 20742}
\newcommand\UNCW{University of North Carolina, Wilmington, North Carolina 28403}
\newcommand\VaTech{Virginia Polytechnic Institute \& State University,
Blacksburg, Virginia 24061}
\newcommand\Yerevan{Yerevan Physics Institute, Yerevan, Armenia}

\begin{document}

\preprint{preprint-ID}

\title{Proton Spin Structure in the Resonance Region}

\author{F.~R.~Wesselmann}	\affiliation{\UVa}\affiliation{\NSU}
\author{K.~Slifer} 		\affiliation{\UVa}
\author{S.~Tajima} 		\affiliation{\UVa}
\author{A.~Aghalaryan}  	\affiliation{\Yerevan}
\author{A.~Ahmidouch}		\affiliation{\NCAT}
\author{R.~Asaturyan\footnote[2]{deceased.}}		\affiliation{\Yerevan}
\author{F.~Bloch}		\affiliation{\UBasel}
\author{W.~Boeglin}		\affiliation{\FIU}
\author{P.~Bosted}		\affiliation{\Jlab}
\author{C.~Carasco}		\affiliation{\UBasel}
\author{R.~Carlini}		\affiliation{\Jlab}
\author{J.~Cha} 		\affiliation{\MissSU}
\author{J.~P.~Chen}		\affiliation{\Jlab}
\author{M.~E.~Christy}		\affiliation{\HamptonU}
\author{L.~Cole}		\affiliation{\HamptonU}
\author{L.~Coman}		\affiliation{\FIU}
\author{D.~Crabb}		\affiliation{\UVa}
\author{S.~Danagoulian} 	\affiliation{\NCAT}
\author{D.~Day} 		\affiliation{\UVa}
\author{J.~Dunne}		\affiliation{\MissSU}
\author{M.~Elaasar}		\affiliation{\SUNO}
\author{R.~Ent} 		\affiliation{\Jlab}
\author{H.~Fenker}		\affiliation{\Jlab}
\author{E.~Frlez}		\affiliation{\UVa}
\author{L.~Gan}		        \affiliation{\UNCW}
\author{D.~Gaskell}		\affiliation{\Jlab}
\author{J.~Gomez}		\affiliation{\Jlab}
\author{B.~Hu}  		\affiliation{\HamptonU}
\author{M.~K.~Jones}		\affiliation{\Jlab}
\author{J.~Jourdan}		\affiliation{\UBasel}
\author{C.~Keith}		\affiliation{\Jlab}
\author{C.~E.~Keppel}		\affiliation{\HamptonU}
\author{M.~Khandaker}		\affiliation{\NSU}
\author{A.~Klein}		\affiliation{\ODU}
\author{L.~Kramer}		\affiliation{\FIU}
\author{Y.~Liang}		\affiliation{\HamptonU}
\author{J.~Lichtenstadt}	\affiliation{\TelAviv}
\author{R.~Lindgren}		\affiliation{\UVa}
\author{D.~Mack}		\affiliation{\Jlab}
\author{P.~McKee}		\affiliation{\UVa}
\author{D.~McNulty}		\affiliation{\UVa}
\author{D.~Meekins}		\affiliation{\Jlab}
\author{H.~Mkrtchyan}		\affiliation{\Yerevan}
\author{R.~Nasseripour} 	\affiliation{\FIU}
\author{I.~Niculescu}		\affiliation{\Jlab}
\author{K.~Normand}		\affiliation{\UBasel}
\author{B.~Norum}		\affiliation{\UVa}
\author{D.~Pocanic}		\affiliation{\UVa}
\author{Y.~Prok}		\affiliation{\UVa}
\author{B.~Raue}		\affiliation{\FIU}
\author{J.~Reinhold}		\affiliation{\FIU}
\author{J.~Roche}		\affiliation{\Jlab}
\author{D.~Rohe}		\affiliation{\UBasel}
\author{O.~A.~Rond\'{o}n}	\affiliation{\UVa}
\author{N.~Savvinov}		\affiliation{\UMD}
\author{B.~Sawatzky}		\affiliation{\UVa}
\author{M.~Seely}		\affiliation{\Jlab}
\author{I.~Sick}		\affiliation{\UBasel}
\author{C.~Smith}		\affiliation{\UVa}
\author{G.~Smith}		\affiliation{\Jlab}
\author{S.~Stepanyan}		\affiliation{\Yerevan}
\author{L.~Tang}		\affiliation{\HamptonU}
\author{G.~Testa}		\affiliation{\UBasel}
\author{W.~Vulcan}		\affiliation{\Jlab}
\author{K.~Wang}		\affiliation{\UVa}
\author{G.~Warren}		\affiliation{\UBasel}\affiliation{\Jlab}
\author{S.~Wood}		\affiliation{\Jlab}
\author{C.~Yan} 		\affiliation{\Jlab}
\author{L.~Yuan}		\affiliation{\HamptonU}
\author{J.~Yun} 		\affiliation{\VaTech}
\author{M.~Zeier}		\affiliation{\UVa}
\author{H.~Zhu} 		\affiliation{\UVa}

\collaboration{Resonance Spin Structure Collaboration}
\noaffiliation

\date{March 30, 2007}

\begin{abstract}
  We have   examined the spin structure of the proton
  in the region of the nucleon resonances 
  ($1.085 \; \mathrm{GeV} < W < 1.910 \; \mathrm{GeV}$)
  at an average four momentum transfer of $Q^2 = 1.3 \; \mathrm{GeV}^2$.
  Using the Jefferson Lab polarized electron beam, 
  a spectrometer, and a  
  polarized solid target, we measured the asymmetries 
  $A_{\|}$ and $A_{\perp}$  to high precision,  
  and extracted the asymmetries $A_1$ and $A_2$,
  and the spin structure functions $g_1$ and $g_2$.
  We found a notably non-zero $A_{\perp}$, 
  significant contributions from higher-twist effects,
  and only weak support for polarized quark--hadron duality.
\end{abstract}


\maketitle

\enlargethispage{11pt}
Ever since the first polarized EMC experiment found that the proton's spin is
not fully carried by its valence quarks~\cite{Ashman:1987hv}, 
the nucleon spin structure has been studied extensively,
for example at SLAC~\cite{e143}, CERN~\cite{Adams:1997hc}, and 
DESY~\cite{Ackerstaff:1997ws}.
The main focus has been on 
kinematics in the deep inelastic scattering (DIS) 
region and with large momentum transfer (high $Q^{2}$) where observations 
can be readily interpreted in a perturbative QCD framework.
In more recent years, lower energy regimes 
($Q^{2} \sim 1 \; \mathrm{GeV}^2$) have grown in importance, 
where the transition from the asymptotically free to the
bound configuration of the quarks can be probed.
Evaluating the requisite moments of the spin structure functions 
requires data or 
model dependent extrapolations, up to parton 
momentum fraction $x = 1$, 
where the electromagnetic scattering probe interacts with 
the proton as a whole,
including the kinematic region dominated by nucleon resonances.

The new experimental focus on the larger $x$ and lower $Q^{2}$ regimes 
so far has been concentrated on longitudinal polarization 
as the dominant, and technically more accessible, component.
This has limited investigations to the spin structure function $g_1$, 
which represents the charge-weighted quark helicity distributions.
Due to the success of DIS interpretations based on the na\"{i}ve parton model, 
which is limited to longitudinal spin components, 
little emphasis has been placed on transverse spin studies.
In contrast, the operator product expansion (OPE) approach to QCD 
includes transverse spin 
starting from leading twist~\cite{Kodaira:1979ib,Jaffe:1989xx} and is 
applicable at all kinematics.
Since each higher order of twist, interpretable as increased correlation 
between partons, adds another $1/Q$ term, higher twist contributions 
should be more prominent at low $Q^2$.
Although transverse spin is suppressed in DIS at leading twist, thus motivating
the twist-2 WW approximation \cite{Wandzura:1977qf}

\begin{equation}
  g_2^\textrm{\tiny{WW}}(x,Q^2) = -g_1(x,Q^2)
            + \int_x^1 \!\! g_1(y,Q^2) \; \frac{dy}{y} ,
  \label{eq:g2ww}
\end{equation}
twist-3 should contribute significantly to $g_2$.
The OPE relates the twist-3 matrix element $d_2$, which represents quark-gluon correlations, 
to the third moments of $g_1$ and $g_2$~\cite{Jaffe:1989xx},
\begin{equation}
 d_2 \,  = \,  3 \int_0^1 \!\!\! x^2 (g_2-g_2^\textrm{\tiny{WW}}) \, dx
     \,  = \,    \int_0^1 \!\!\! x^2 (2 g_1 + 3 g_2) \, dx   ,
 \label{eq:d2}
\end{equation}
permitting comparison with QCD lattice calculations, and thus 
providing a clean test of the theory.

Also, the phenomenon of quark--hadron duality~\cite{BloomGilman} 
has captured much interest
\cite{Airapetian:2002rw,Bianchi:2003hi,Melnitchouk:2005zr,Dong:2006yb,Bosted:2006gp}.
This concept connects the perturbative QCD description of 
quarks and gluons to the hadronic description at lower energies.
Whereas global duality relates the entire resonance region to extrapolations 
of the asymptotic structure functions, local duality relates the extrapolated 
quantities to restricted resonance regions.
Both have been observed in unpolarized scattering~\cite{Niculescu:2000tk},
but precision measurements of the spin structure functions in the resonance 
region are needed to determine whether they also display duality.
If local duality, as first seen by Bloom and Gilman~\cite{BloomGilman}, 
were observed for $g_1$, this would not only demonstrate 
its universal, rather than accidental nature \cite{Close:2001ha},
but it could also serve to justify extrapolations.
The potential usefulness of the latter for the determination of the
moments increases at low $Q^2$, where the resonances extend over 
wider ranges of $x$ than at high $Q^2$.

Experiment E01-006 was conducted in Hall~C of 
the Thomas Jefferson National Accelerator Facility
by the Resonance Spin Structure (RSS) collaboration.
Utilizing established procedures and equipment, we have measured 
the asymmetries 
$A_{\parallel}$ and $A_{\perp}$ in the scattering of polarized electrons off 
a polarized proton target.
These asymmetries are defined as the dimensionless, relative difference between
the cross sections with 
parallel and anti-parallel (or perpendicular and anti-perpendicular) alignment
of the proton and the electron spins.
Focusing exclusively on the hadronic vertex, we are left with 
the virtual photon asymmetries
$A_1$ and $A_2$, which are functions of the virtual photon's 
four-momentum-squared $-Q^2$ and the invariant mass $W$ of the final state.
Without a measurement of $A_{\perp}$, this step would require the 
use of a model or a sweeping assumption.
Further, switching from an external scattering view 
to an internal structure interpretation, we can obtain the 
spin structure functions $g_1(x,Q^2)$ and $g_2(x,Q^2)$, where $Q^2$ is 
interpretable as the energy scale set by the probe and, in DIS, 
$x$ represents the fraction of the proton momentum carried by the 
constituent the probe interacted with.
The exact relation between these quantities, and other relevant
definitions, can be found in Sec.~II of Ref.~\cite{e143} or in
Sec.~2.1 of Ref.~\cite{Melnitchouk:2005zr}.



The experiment used Jefferson Lab's
continuous, polarized electron beam with energy of 
$5.755 \; \mathrm{GeV}$ and a nominal current of $100 \; \mathrm{nA}$. 
The beam polarization was measured by a M{\o}ller 
polarimeter~\cite{Hauger:1999iv} installed upstream of the target, and 
the beam helicity was flipped at $30 \; \mathrm{Hz}$ on a pseudo-random basis.

Frozen $^{15}\mathrm{NH}_{3}$, in $1 \! - \! 2 \; \mathrm{mm}$ fragments, 
was used as the proton target in the University of Virginia 
apparatus \cite{Crabb:1995xi} in which 
a $^4\mathrm{He}$ evaporation refrigerator at $1 \; \mathrm{K}$ coupled with a $5 \; \mathrm{T}$ 
polarizing magnet created a stable polarization environment.
The polarization population enhancement was achieved via 
microwave pumping (dynamic nuclear polarization) 
and measured by an NMR system using pickup coils embedded in the 
target material. 
For the $A_{\perp}$  measurement, the entire target apparatus was rotated 
by $90^{\circ}$ in the scattering plane.
About equal amounts of data were taken with each polarization direction, 
flipping the nuclear spins by adjusting the microwave frequency, to 
reduce systematic effects.
To maintain uniform polarization in the bulk target material, the beam was
continually moved across the face of the target in a 
$1 \; \mathrm{cm}$ maximum radius 
spiral raster pattern around the beam axis.
This extra degree of freedom required a dedicated beam position 
monitor~\cite{Steinacher:2000xr} for accurate event reconstruction.

Scattered electrons were detected using the High Momentum Spectrometer (HMS), 
positioned at a scattering angle of $13.15^{\circ}$.
Two different HMS momentum settings were used,
$4.078$ and $4.703 \;\mathrm{GeV}$, to cover the desired wide 
range in $W$, resulting in $0.8 < Q^{2} < 1.4 \; \mathrm{GeV}^2$.
A detector package consisting of hodoscope planes, wire chambers, a 
gas \v{C}erenkov counter, and a lead glass calorimeter allowed for
particle identification and measurement of the event kinematics.
A more detailed description of the apparatus and technique can be found 
in Ref.~\cite{Jones:2006kf}.


Approximately 160 million scattering events were recorded on the proton target,
resulting in highly precise determinations of the parallel and perpendicular
asymmetries.
These are obtained from observed raw event
counting asymmetries which are scaled to $100 \%$ polarization and
corrected for contamination from radiative and dilution processes:
\begin{equation}
  A_{\parallel,\perp} = \frac{1}{f C_N P_b P_t f_{RC}}
    \times \frac{N^- - N^+}{N^- + N^+}
    + A_{RC}
\label{eq:AsyEx}
\end{equation}

Here, $N^\pm$ is the charge corrected observed yield for the
parallel (perpendicular) and anti-parallel (anti-perpendicular) 
spin alignment, respectively.  $P_b$ and $P_t$ are the beam and
target polarizations, $f$ is the dilution factor, $C_N$ is a small 
$^{15}$N nuclear polarization correction, and $f_{RC}$ and $A_{RC}$ are
radiative corrections.

\begin{figure}[ht]
  \center
  \includegraphics[width=0.495\textwidth,keepaspectratio]{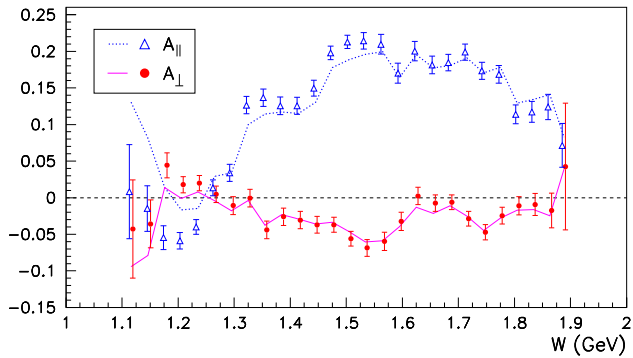}
 \caption{Our measured asymmetries $A_\parallel$ and $A_\perp$,
          fully corrected (points) 
          and without radiative corrections (curves).}
 \label{pAsy}
\end{figure}

\begin{table}[hb]
  \caption{Averaged systematic errors in the asymmetries.}
  \label{Tsyst}
  \center
  \begin{ruledtabular}
   \begin{tabular}{lrr}
	Error Source              & \multicolumn{1}{c}{$A_\parallel$}   
				  & \multicolumn{1}{c}{$A_\perp$}     \\ \hline
	Target Polarization  	  & 	       &   2.9 \%	          \\
	Beam Polarization   	  &  \raisebox{6pt}[0pt][0pt]{$\left\}\rule{0pt}{8pt}\right.$1.1 \%}
					      &   1.3 \%		 \\
	Dilution Factor  	  &  4.9 \%   &   4.9 \%		\\
	Radiative Corrections  	  &  2.7 \%   &  12.9 \%		\\
	Kinematic Reconstruction  &  0.4 \%   &   0.4 \%		\\
   \end{tabular}
  \end{ruledtabular}
\end{table}

The corrected asymmetries $A_\parallel$ and $A_\perp$ are
shown in Fig.~\ref{pAsy};
$A_\perp$ is notably non-zero.
The average proton polarization was $62 \%$ ($70 \%$), and the beam polarization
was $71 \%$ ($66 \%$) during the parallel (perpendicular) running.
For the parallel alignment, the product $P_b \times P_t$ 
was derived by normalizing the measured
elastic asymmetry \cite{Jones:2006kf}
to the known value, resulting in better accuracy than achievable from
direct measurements.
In the perpendicular case, the limited knowledge 
of the elastic asymmetry made the direct measurement 
of $P_b$ and $P_t$ the better choice.
The systematic errors are summarized in Table~\ref{Tsyst},
highlighting the lack of models and data for perpendicular radiative corrections.

The dilution factor represents the fraction of events 
that truly scattered from a polarized proton in the target.
It was determined from the ratio of free proton to total target rates 
calculated via a Monte Carlo simulation which had been matched to 
calibration data acquired specifically for this purpose.
The QFS parameterization \cite{QFS},
modified to improve agreement with our data, 
was used as input for the Born inelastic cross sections for
$\mathrm{A} \geq 3$ nuclei. 
Fits to Hall~C inelastic {\it e--p} 
data \cite{Liang:2004tj} were used for the H contribution.
The unpolarized structure function $F_1$ and the ratio $R$ of the
longitudinal to transverse cross sections are derived from the same 
fits~\cite{christy_F2R,Liang:2004tj}.
The uncertainty in these cross sections was the dominant source of 
systematic error for the dilution correction.

Convoluting radiative prescriptions with models of the resonance region, 
the elastic peak, and our target, we obtained radiated cross sections 
and asymmetries.
The external radiative corrections were determined using the procedure 
established in~\cite{Stein:1975yy}, 
while the POLRAD software \cite{Akushevich:1997di}
was used to determine the internal radiative corrections.
The resonance fit model was iteratively improved, until the radiated 
values matched our experimental data.
The model then trivially provided the corrections to 
our measurements, with $f_{RC}$ accounting for the radiative dilution 
from the elastic tail and $A_{RC}$ for all other influences.

\begin{figure}[b]
  \center
  \includegraphics[width=0.495\textwidth,keepaspectratio]{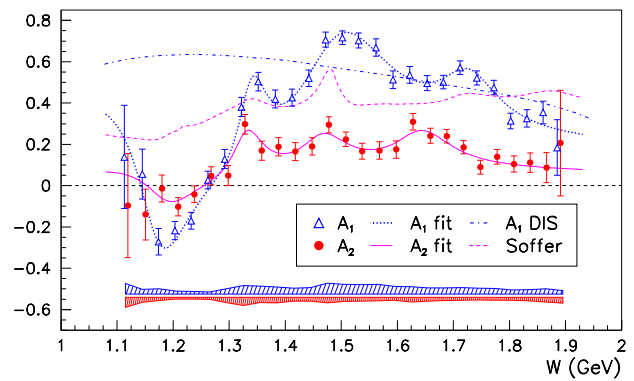}
  \caption{Virtual photon asymmetries $A_1$ and $A_2$ from our data 
           and corresponding fits.
	   Also shown is the E155 fit to DIS data~\cite{Anthony:2000fn,Wesselmann:2000xu}, 
	   evaluated at our $(x,Q^2)$, 
	   and the Soffer limit for $A_2$ \cite{Soffer:2000zd}, based on our
	   $A_1$ fit.
	   The upper error band indicates the systematic error in $A_1$, the lower one $A_2$.}
 \label{pA12}
\end{figure}

\begin{figure}[t]
    \center
    \includegraphics[width=0.495\textwidth,keepaspectratio]{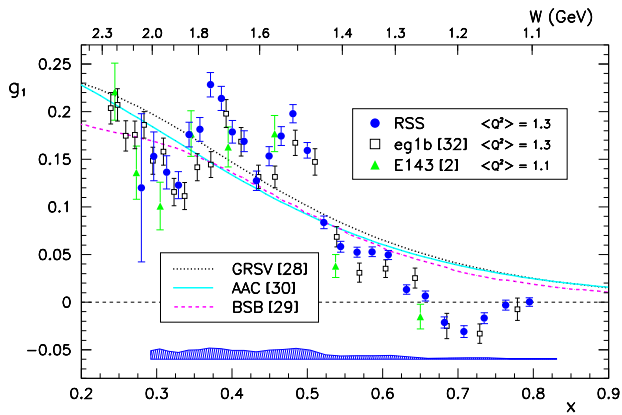}
    \caption{Results for $g_1$ from this experiment (RSS) and other relevant 
             data \cite{e143,Dharmawardane:2006zd},
	     as well as target mass corrected NLO PDFs.
	     The upper scale shows $W$ (at $Q^2 = 1.3 \; \mathrm{GeV}^2$) 
	     for reference.}
    \label{pg1}
  \end{figure}
\begin{figure}[t]
    \center
    \includegraphics[width=0.495\textwidth,keepaspectratio]{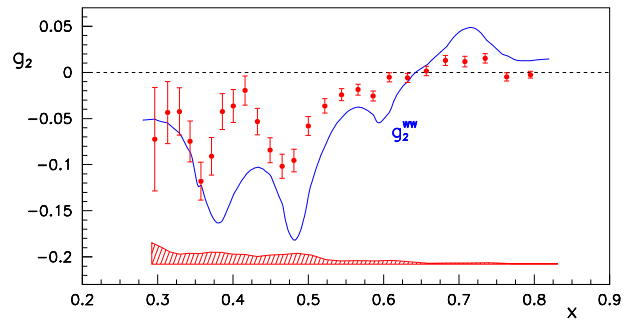}
   \caption{Our (RSS) values for $g_2$ and the approximation 
	 $g_2^\textrm{\tiny{WW}}$ (Eq.~\ref{eq:g2ww})
	 as evaluated from our data.}
   \label{pg2}
  \end{figure}

We extract the virtual photon asymmetries 
$A_1$ and $A_2$, shown in Fig.~\ref{pA12}, 
from the corrected physics asymmetries $A_{||}$ and $A_{\perp}$, 
using only $R$ as model dependent input.
The spin structure functions $g_1$ and $g_2$ 
(Figs. \ref{pg1} and~\ref{pg2})
are then obtained utilizing $F_1$.
The uncertainties in $F_1$ and $R$ are 
included in our total systematic error
and in the error bands of our plots.

We have fitted the $W$ dependence of our $A_1$ and $A_2$ data using an approach 
similar to that applied to unpolarized cross sections 
in Ref.~\cite{Stein:1975yy}, substituting for the DIS component a form based on 
the phenomenological spin structure parameterizations of 
Refs.~\cite{Anthony:2000fn,Wesselmann:2000xu}. 
These fits served as input in the iterative procedure to obtain our radiative 
corrections and to calculate the integrals of $g_1$ and $g_2$ at constant $Q^2$.
Each spin asymmetry was fitted independently, since they represent different 
physical quantities.

To test quantitatively for global duality in $g_1$, we can integrate 
in $x$ over the resonance region and compare the results obtained from
resonance data and DIS extrapolations (Fig.~\ref{pg1}).
We used our fit to integrate over the resonance region 
($1.09 < W < 1.91 \; \mathrm{GeV}$) and took the average of the 
integrals from several DIS extrapolations calculated from 
target mass corrected~\cite{Blumlein:1998nv}, next-to-leading order
parton distribution functions (NLO PDFs)
\cite{Gluck:1995yr,Goto:1999by,Bourrely:2005kw} 
over the same range of $W$ at our average $Q^2 = 1.3 \; \mathrm{GeV}^2$.
We found 
the ratio of integrals, PDFs to data, to be $1.17 \pm 0.08$,
indicating agreement at only the two sigma level, and 
suggesting that PDF extrapolations into 
the resonance region may not be valid at this $Q^2$.
The ratios for restricted, but still rather broad, $W$ ranges 
differ from unity by several sigmas, demonstrating that 
\emph{local} polarized duality is not valid at our $Q^2$:
$1.09 < W <1.4 \; \mathrm{GeV}$ is $6.47 \pm 0.95$,
and 
$1.4 < W < 1.91 \; \mathrm{GeV}$ is $0.87 \pm 0.06$.
Including also large $x$ resummations for the PDFs~\cite{Bianchi:2003hi}, 
the global ratio changed to $1.42 \pm 0.10$.
The quoted errors are based on the data integrals only,
including a $0.4 \%$ contribution from computing our fit at fixed $Q^2$.
Our results are in good agreement with the recent results from
CLAS~\cite{Bosted:2006gp}.

Approximate global duality within errors was reported by 
\cite{Airapetian:2002rw}, based on 
$A_1$ resonance data averaged over a broad $Q^2$ range from 
$1.6$ to $2.9 \; \mathrm{GeV}^2$ and compared to a DIS fit to data. 
The weak $Q^2$ dependence of $A_1$ (within large errors) allows for 
averaging, instead of calculating the ratio at each $Q^2$ value 
as is required for testing duality in the structure functions. 
But duality in the spin asymmetry $A_1 \propto g_1 / F_1$ could be due to
accidental cancellations in the ratio $g_1 / F_1$.


Our results for $g_2$ are much clearer, especially 
in the framework of the QCD OPE. 
The comparison of our data and the $g_2^\textrm{\tiny{WW}}$ approximation, 
evaluated from our measurements of $g_1$,
provides strong evidence of the significance of higher-twist terms
at this $Q^2$, as shown in Fig.~\ref{pg2}.
Combining our measurements of $g_1$ and $g_2$, we can investigate specifically 
the twist-3 contribution via the matrix element 
$d_2$ (Eq.~\ref{eq:d2}).
Over the measured range ($0.29 \! < \! x \! < \! 0.84$), we find 
$\overline{d_2} = 0.0057 \pm 0.0009 \; \mathrm{(stat)} 
                         \pm 0.0007 \; \mathrm{(syst)}$,
including a $4 \%$ contribution to the systematic error from our 
fit's assumed $Q^2$ dependence.
This significantly non-zero result highlights the limitation of 
leading-twist approximations.
Extrapolating this result to $Q^2 = 5 \; \mathrm{GeV}^2$,
assuming a $1/Q$ dependence,
we find $\overline{d_2} = 0.0029$
compared to the SLAC result 
$d_2 = 0.0032 \pm 0.0017$~\cite{Anthony:2002hy}.

In summary, our results 
significantly increase the available information 
on the proton spin structure:  
These new data provide a connection to the measurements at DIS kinematics
and fill a significant void in the explored regions.
Our measurement with transverse spin arrangement is
the first in the resonance region, with notably non-zero results.
Our data clearly indicate the importance of 
higher twist contributions and thus quark--gluon correlations.
We have established that Bloom-Gilman polarized duality is 
meaningful only
for the resonance region as a whole, although local polarized duality
may yet be observed at higher $Q^2$ ranges.

We would like to thank the Hall~C technical staff and
the accelerator operators for their efforts and dedication.
This work was supported by 
the Department of Energy, 
the National Science Foundation, 
the Schweizerische Nationalfonds,
and by the 
Institute of Nuclear and Particle Physics of the University of Virginia. 
The Southern Universities Research Association
(SURA) operates the Thomas Jefferson National Accelerator Facility 
under contract for the 
United States Department of Energy.

\clearpage

\appendix

\section{Appendix}

The results of our fits to $A_1$ and $A_2$ are detailed in table~\ref{TfitA12} 
and plotted against our data in figure~\ref{pA12}.
They are based on the method of Ref.~\cite{Stein:1975yy},
combining a number of Breit-Wigner shaped resonances $\mathit{B\!W}_{\!\!i}$
and a deep-inelastic tail:

\begin{equation*}
   \mathrm{fit}   =    \underbrace{ \sum _{i=1}^{4} \mathit{B\!W}_{\!\!i} }_{\mathrm{Resonances}}
                 \ + \ \underbrace{ x^\alpha \sum _{n=0}^{3} \beta_n \; x^n }_{\mathrm{DIS}}
                 \ \underbrace{ \times \ \frac{1}{\sqrt{Q^2} } \raisebox{0pt}[0pt][12.9pt]{\;} }_{A_2 \ \mathrm{only}}
\end{equation*}

where

\begin{equation*}
   \mathit{B\!W}_{\!\!i} = \frac{ a_i \ \kappa_i^{2} \ w_i^2 \ \Gamma_i \ \Gamma^\gamma_i
                       }{ \kappa_{cm}^{2} \ \left[   ( w_i^2 - W^2 )^2 
		                                  + w_i^2 \ \Gamma_i^{2} \right] }
\end{equation*}

and

\begin{eqnarray*}
	      \Gamma _i & = & g_i 
	                      \left( \frac{q_{cm}}{q_i} \right)^{(2 l_i + 1)} 
                              \left( \frac{ q_i^2    + X_i^2
			        	 }{ q_{cm}^2 + X_i^2} \right)^{l_i}
   \\
        \Gamma^\gamma_i & = & g_i 
                              \left( \frac{\kappa_{cm}}{\kappa_i}     \right)^{(2 j_i)}  \ \ 
                              \left( \frac{\kappa_i^{2}   + X_i^2
			   		  }{\kappa_{cm}^2 + X_i^2} \right)^{j_i}
\end{eqnarray*}

with

\begin{eqnarray*}
        \kappa_i    & = & \sqrt{ \frac{ ( w_i^2 + M^2 + Q^2 )^2 }{ 4 w_i^2 } - M^2}
    \\
        q_i	    & = & \sqrt{ \frac{ ( w_i^2 + M^2 - m_{\pi}^2 )^2 }{ 4 w_i^2 } - M^2}
    \\
        \kappa_{cm} & = & \sqrt{ \frac{ ( W^2 + M^2 + Q^2 )^2 }{ 4 W^2 } - M^2}
    \\
        q_{cm}      & = & \sqrt{ \frac{ ( W^2 + M^2 - m_{\pi}^2 )^2 }{ 4 W^2 } - M^2}
\end{eqnarray*}

\begin{table}[htb]
  \caption{Results of Fits to $A_1$ and $A_2$, 
           each based on 28 data points as plotted in figure~\ref{pA12}.}
  \label{TfitA12}
  \center
  \begin{ruledtabular}
   \begin{tabular}{l
		   D{.}{.}{3}@{$\;\pm$\!\!\!\!\!\!\!\!\!\!\!\!\!\!\!\!\!\!\!\!}D{.}{.}{3}
		   D{.}{.}{3}@{$\;\pm$\!\!\!\!\!\!\!\!\!\!\!\!\!\!\!\!\!\!\!\!}D{.}{.}{3}}
               \multicolumn{1}{r}{Parameter} 
	     & \multicolumn{2}{c}{$A_1$ Fit} 
	     & \multicolumn{2}{c}{$A_2$ Fit} \\ \hline
    $\chi^2 / \mathrm{dof}$  &  \multicolumn{2}{c}{1.24}  &  \multicolumn{2}{c}{1.42}  \\ \hline
   	  $a_1$     &	-0.562 & 0.133    &    -0.147 & 0.052	    \\        
   	  $a_2$     &	 0.401 & 0.069    &	0.220 & 0.049	    \\        
   	  $a_3$     &	 0.537 & 0.042    &	0.158 & 0.029	    \\        
   	  $a_4$     &	 0.246 & 0.044    &	0.173 & 0.029	    \\ \hline 
   	  $w_1$     &	 1.204 & 0.010    &	1.214 & 0.021	    \\        
   	  $w_2$     &	 1.345 & 0.005    &	1.338 & 0.010	    \\        
   	  $w_3$     &	 1.544 & 0.012    &	1.479 & 0.021	    \\        
   	  $w_4$     &	 1.734 & 0.009    &	1.653 & 0.015	    \\ \hline 
   	  $g_1$     &	 0.161 & 0.065    &	0.137 & 0.086	    \\        
   	  $g_2$     &	 0.063 & 0.025    &	0.071 & 0.035	    \\        
   	  $g_3$     &	 0.283 & 0.058    &	0.113 & 0.082	    \\        
   	  $g_4$     &	 0.125 & 0.043    &	0.129 & 0.062	    \\ \hline 
   	  $\alpha$  &	 0.031 & 0.391    &	0.458 & 0.375	    \\ \hline 
   	  $\beta_0$ &	 0.186 & 0.074    &	0.100 & 0.040	    \\        
   	  $\beta_1$ &	-0.032 & 0.158    &	0.094 & 0.079	    \\        
   	  $\beta_2$ &	-0.393 & 0.272    &    -0.119 & 0.133	    \\        
   	  $\beta_3$ &	 0.957 & 0.403    &    -0.025 & 0.202		      
   \end{tabular}
  \end{ruledtabular}
\end{table}

We have optimized the values of amplitude $a_i$, centroid $w_i$, 
and width $g_i$ of four Breit-Wigner resonances,
as well as the exponent $\alpha$ and the polynomial coefficients
$\beta_n$ for the DIS tail.
The values of the parameters $X_i$,
$l_i$, and $j_i$ were kept as in Ref.~\cite{Stein:1975yy}.
While the fits are a function of $W$, the DIS contribution
is evaluated at the mean values of $x$ and $Q^2$ 
of our data at that $W$~bin.

In addition to the explicit contribution in the terms $\kappa_i$ 
and $\kappa_{cm}$, and the indicated $1/\sqrt{Q^2}$ term for the DIS part of $A_2$,
our fit function depends on $Q^2$ via the 
interrelation of $W$, $x$ and $Q^2$:
\[ W^2 = M^2 - Q^2 + \frac{Q^2}{x} \]
The resonance terms are otherwise $Q^2$-independent.

The choice of four resonances resulted in better fits than using only three,
an option also supported by examining the second derivatives of the data smoothed 
with a spline curve.
Additionally, the fits as reported here also gave excellent agreement 
with the cross sections
$\sigma_{\mathrm{TT'}}$ and $\sigma_{\mathrm{LT'}}$ obtained from our data and a
$\sigma_{\mathrm{T}}$ model~\cite{christy_F2R}.

Table~\ref{Tpdfratios} details the results of duality integral ratios 
for various sub-ranges (those matching the resonances assumes in our fit 
to $A_1$, as well as those of \cite{Melnitchouk:2005zr}) and for the different 
PDFs \cite{Gluck:1995yr,Goto:1999by,Bourrely:2005kw}.
The PDF integrals over individual resonances differed
significantly from the data value, indicating that at our $Q^2$,  
local duality is not observed in $g_1^p$.

\begin{table}[htb]
  \caption{Ratio of NLO PDFs to data,
           integrated over the indicated ranges.
	   The top five $W$ ranges are based on the location of the Breit-Wigner
	   shapes resulting from our fit to $A_1$ and
	   the last two match those of Ref.~\cite{Melnitchouk:2005zr};
	   all are bounded by our acceptance.
	   The elastic contribution \cite{Melnitchouk:2005zr} was calculated 
	   using the dipole form of the form factors.}
  \label{Tpdfratios}
  \center
  \begin{ruledtabular}
   \begin{tabular}{l
                   D{.}{.}{2}@{\,-\!\!}D{.}{.}{2}@{\!\!}l
		   D{.}{.}{2}
		   D{.}{.}{2}
		   D{.}{.}{2}
		   D{.}{.}{2}@{$\pm$\!\!}D{.}{.}{2}}
                 & \multicolumn{3}{c}{$W$ Range} 
	         & \multicolumn{1}{c}{BSB} 
	         & \multicolumn{1}{c}{GRSV} 
	         & \multicolumn{1}{c}{AAC} 
	         & \multicolumn{2}{c}{Average} \\ \hline
      Delta      &  1.09 & 1.30 & GeV  &  3.41  &  4.18  &  3.96  &  3.93 & 0.58 \\
      R1         &  1.30 & 1.39 & GeV  &  1.28  &  1.44  &  1.33  &  1.36 & 0.10 \\
      R2         &  1.39 & 1.68 & GeV  &  0.77  &  0.82  &  0.75  &  0.78 & 0.05 \\
      R3         &  1.68 & 1.79 & GeV  &  0.77  &  0.84  &  0.77  &  0.79 & 0.06 \\
       \hline
      Global     &  1.09 & 1.91 & GeV  &  1.11  &  1.23  &  1.14  &  1.17 & 0.08 \\
       \hline
      Delta + R1 &  1.09 & 1.40 & GeV  &  5.78  &  6.84  &  6.44  &  6.47 & 0.95 \\
      R2 +       &  1.40 & 1.91 & GeV  &  0.84  &  0.91  &  0.84  &  0.87 & 0.06 \\
   \end{tabular}
  \end{ruledtabular}
\end{table}

\clearpage

\end{document}